\documentclass[aps,pra,twocolumn,superscriptaddress]{revtex4-2}

\usepackage{graphicx}
\usepackage{dcolumn}
\usepackage{bm}
\usepackage{amsmath}
\usepackage{amssymb}
\usepackage{latexsym}
\usepackage{epsfig}
\usepackage{amsbsy}
\usepackage{array}
\usepackage{amssymb}
\usepackage{setspace}
\usepackage{bm}
\usepackage{epstopdf}
\usepackage{subfigure}
\usepackage{subfigure}
\usepackage{color}

\def\sint{\ifmmode{- \!\!\!\!\!\! \int}
    \else{\hbox{$- \!\!\!\! \int \ $}}\fi}

\begin{document}

\title{Theoretical analysis towards accurate optomechanical detection of quantum gravity effects}
\author{Ying Li}
\affiliation{College of Sciences, Northeastern University, Shenyang 110819, China}
\author{Yan Li}
\affiliation{School of Physics and Information Technology, Shaanxi Normal University, Xi’an, 710062, China}
\affiliation{Department of Physics, The Chinese University of Hong Kong, Shatin, New Territories, Hong Kong, China}
\author{Chengsong Zhao}
\affiliation{College of Sciences, Northeastern University, Shenyang 110819, China}
\affiliation{Research Center of Quantum Physics, Northeastern University, Shenyang 110819, China}
\author{Najmeh Eshaqi-Sani}
\affiliation{ICMM-CSIC, Sor Juana Inés de la Cruz 3, 28049 Madrid, Spain}
\author{Wenlin Li}
\email{liwenlin@mail.neu.edu.cn}
\affiliation{College of Sciences, Northeastern University, Shenyang 110819, China}
\affiliation{Research Center of Quantum Physics, Northeastern University, Shenyang 110819, China}

\date{\today}
\begin{abstract}
Optomechanical systems offer a promising platform for observing dynamical signatures of quantum gravity through precision measurements of quantum harmonic oscillator dynamics. However, most existing analyses consider only the linear radiation-pressure interaction while neglecting higher-order optomechanical couplings and laser phase noise. These neglected contributions can be comparable in magnitude to the predicted quantum-gravity corrections and may therefore introduce spurious signals or mask the genuine physical effect. Here we reanalyze two experimentally realized platforms, a Fabry–Pérot optomechanical system and a membrane-in-the-middle optomechanical system, by incorporating the complete nonlinear dynamics and realistic laser phase noise. Using measured device parameters, we derive revised protocols for generalized uncertainty principle tests and establish practical sensitivity bounds. Our results demonstrate that previous idealized estimates significantly overestimate the achievable resolution, underscoring the necessity of including higher-order interactions and implementing effective laser phase noise suppression in realistic assessments of optomechanical quantum gravity tests.
\end{abstract} 
\maketitle

\section{Introduction}
Although physicists remain divided on the appropriate theoretical framework for describing gravity in regimes where quantum effects are significant, many candidate theories of quantum gravity predict the existence of a fundamental minimum length scale at the Planck scale~\cite{Adler1982,Garay1995,Hossenfelder2013,Amelino-Camelia2002}. In the weak-gravity limit, such a theory is expected to reduce to a modified form of quantum mechanics. In this regime, the uncertainty relations would require corrections compared with those of standard quantum mechanics, since the Heisenberg uncertainty principle permits arbitrarily fine spatial resolution provided that sufficient momentum uncertainty is accepted~\cite{Pikovski2012}. In 1990s, Refs.~\cite{Amati1987,Gross1988,Maggiore1993,Scardigli1999,Jizba2010} proposed a modified form of the uncertainty principle, known as the generalized uncertainty principle (GUP):
\begin{equation}
\begin{split}
\Delta X\Delta P\geq\dfrac{\hbar}{2}\left[1+\beta_0\left(\dfrac{L_p\Delta P}{\hbar}\right)^2\right].
\end{split}
\label{eq:GUP}
\end{equation}
The above equation implies a minimum resolvable length scale $\sqrt{\beta_0}L_p$. In fact, the dimensionless parameter $\beta_0$ encodes quantitative information about the underlying quantum gravity theory. Currently, it is known only that $1\leq\beta_0<10^{34}$, a range that corresponds to energies up to the electroweak scale~\cite{Pikovski2012,Das2008,Marin2013}. Were $\beta_0$ to exceed the upper bound of this interval, signatures of the GUP would already be observable in electroweak-scale experiments. Beyond this phenomenological constraint, no theoretical framework can predict the precise value of $\beta_0$; its determination therefore necessitates dedicated experimental investigation. At present, direct measurements of generalized uncertainty relations, such as those probing the associated zero-point energy, can experimentally constrain the GUP parameter to $\beta_0<10^{33.4}$~\cite{Marin2013}. Moreover, the majority of proposed measurement schemes instead rely on reformulating Eq.~\eqref{eq:GUP} in terms of the modified commutator~\cite{Pikovski2012,Bawaj2015,Bonaldi2020}:
\begin{equation}
\begin{split}
[\hat{X},\hat{P}]=i\hbar\left[1+\beta_0\left(\dfrac{L_p\hat{P}}{\hbar}\right)^2\right],
\end{split}
\label{eq:GUPc}
\end{equation}
These approaches have been employed to constrain the GUP parameter through a range of physical systems and phenomena, including the ground-state Lamb shift~\cite{Das2008,Ali2011} and the 1S–2S transition frequency in hydrogen~\cite{Quesne2010}, tests of the equivalence principle~\cite{Ghosh2014}, and observations of gravitational-wave events. Among these schemes, measurements of the GUP corrections to the  dynamics of a harmonic oscillator have been widely discussed in recent decade, owing to their potential for achieving high precision~\cite{Pikovski2012,Bawaj2015,Bushev2019,Bonaldi2020,Cui2021,Campbell2023,Li2025,Li2026,Li20262}. Moreover, the relative ease of laboratory implementation renders the scheme amenable to further optimization, for instance, by cooling the oscillator to near its quantum ground state~\cite{Li2025} or by enhancing the measurement resolution through the exploitation of dark modes~\cite{Li2026}, nonlinearity~\cite{Li20262}, and related techniques.

In general, the manipulation and measurement of mechanical oscillators are realized within the framework of optomechanics (OMS)~\cite{Pikovski2012,Bonaldi2020,Cui2021,Li2025,Li2026,Li20262}. Previous theoretical models have typically assumed purely single-mode coherent driving of the cavity and treated the field–oscillator interaction as a first-order radiation-pressure coupling~\cite{Aspelmeyer2014}. Neither the noise inherent to real physical systems nor higher-order field–oscillator interactions were incorporated into these models. In virtually all earlier studies, such contributions could be safely neglected because they were smaller than the leading-order interaction by at least five orders of magnitude~\cite{Gao2015,Li20252}. However, when the target signals are the minute corrections predicted by the GUP, this assumption can no longer be made a priori. Because the GUP corrections themselves are of comparable magnitude~\cite{Bawaj2015}, it is necessary to determine whether these previously neglected effects remain negligible and, if not, to include them fully in the analysis.

In this paper, we reanalyze a representative class of OMS schemes for probing GUP effects. The systems we consider incorporate nonlinear dynamics to all orders~\cite{Gao2015,Li20252}, and include laser phase noise~\cite{Abdi2011}. Higher-order interactions can generate false-positive signals that mimic GUP corrections, while phase noise can overwhelm the true mechanical signal. On this basis, we quantitatively reevaluate the performance of such systems, with particular emphasis on their ability to detect GUP corrections in the quantum regime of a harmonic oscillator. Specifically, we examine two experimentally realized OMSs, a Fabry–Pérot (FP) cavity OMS and a membrane-in-the-middle (MiM) OMS. By incorporating the measured parameters of these systems, we rederive the GUP measurement protocol. In contrast to the analyses reported in Refs.~\cite{Bonaldi2020,Li2025,Li2026,Li20262}, which consider only the resolution achievable under ideal conditions, our work establishes a resolution that more accurately reflects the practical limitations imposed by the experimental equipment.

The remainder of this paper is organized as follows. In Sec.~\ref{Probing GUP corrections with resonator dynamics} we review schemes that employ optomechanical systems to measure the GUP. Sec.~\ref{The model and the Langevin equation} presents a comprehensive analysis of the nonlinear dynamics of both FP OMS and MiM OMS, incorporating radiation-pressure interactions to all orders as well as the effects of laser phase noise. In Sec.~\ref{Measuring GUP in the quantum regime}, we give a quantitative precision analysis of the GUP measurement scheme, with particular emphasis on the influence of higher-order nonlinearities and phase noise on the achievable resolution. A summary of our principal results and an outlook for future investigations are provided in Sec.~\ref{Discussion of other situations}.

\section{Probing GUP corrections with resonator dynamics}
\label{Probing GUP corrections with resonator dynamics}
Extending the framework of Ref.~\cite{Bawaj2015}, the present work focuses on the precise characterization of the nonlinear dynamics of a harmonic oscillator, thereby enabling a quantitative extraction of the GUP parameter $\beta_0$. The analysis rests on two central assumptions~\cite{Bawaj2015,Pikovski2012}: First, the harmonic oscillator is assumed to retain its conventional Hamiltonian. For a resonator of mass $m$ and resonance frequency $\omega_b$, it takes the form $H=m\omega_b^2\hat{X}^2/2+\hat{P}^2/2m$; Second, the Heisenberg equations of motion are taken to retain their standard form: for any operator $\hat{o}$, $d\hat{o}/dt=[\hat{o},H]/i\hbar$. Here we introduce the standard dimensionless quadratures $q$ and $\hat{\tilde{p}}$, defined according to $\hat{X}=\sqrt{\hbar/(m\omega_b)}\hat{q}:=x_{\rm ZPF}\hat{q}$ and $P=\sqrt{\hbar m\omega_b}\hat{\tilde{p}}$. Following this transformation, the Hamiltonian reduces to 
\begin{equation}
\begin{split}
H_{m}/\hbar=\dfrac{\omega_b}{2}\left(q^2+\tilde{p}^2\right),
\end{split}
\label{eq:Hamilton_b}
\end{equation}
and the GUP commutator described by Eq.~\eqref{eq:GUPc} then takes the form:
\begin{equation}
\begin{split}
[\hat{q},\hat{\tilde{p}}]=i\left(1+\beta_{\rm NL}\hat{\tilde{p}}^2\right),
\end{split}
\label{eq:GUPcs}
\end{equation}
where $\beta_{\rm NL}=\beta_0(\hbar m\omega_b/m_p^2c^2)$ is the dimensionless GUP parameter. Here, $m_p$ denotes the Planck mass and $c$ is the speed of light in vacuum. To restore the canonical commutation relation $[\hat{q},\hat{p}]=i$, we introduce the auxiliary momentum operator
\begin{equation}
\begin{split}
\hat{\tilde{p}}=\left(1+\beta_{\rm NL}\dfrac{\hat{p}^2}{3}\right)\hat{p}.
\end{split}
\label{eq:auxiliary momentum}
\end{equation}
To first order in $\beta_{\rm NL}$, the Hamiltonian of the mechanical oscillator acquires an additional nonlinear correction quartic in the momentum $  \hat{p}  $, namely
\begin{equation}
\begin{split}
H_{m}/\hbar=\dfrac{\omega_b}{2}\left(\hat{q}^2+\hat{p}^2\right)+\dfrac{1}{3}\omega_b\beta_{\rm NL}\hat{p}^4.
\end{split}
\label{eq:non_Hamilton_b}
\end{equation}
In the regime $\beta_{\rm NL}\ll 1  $, the quartic correction modifies the dynamical oscillation frequency of the coordinate $\hat{q}$, treating the motion perturbatively (or semiclassically), the solution takes the form $\hat{q}(t)=A\sin(\omega_{\rm eff}t+\phi)/\sqrt{2}$ with an effective frequency~\cite{Strogatz1994}:
\begin{equation}
\begin{split}
\omega_{\rm eff}=\omega_b\left(1+\beta_{\rm NL}{A^2}\right):=\omega_b\left(1+\beta_{\rm NL}I\right),
\end{split}
\label{eq:eff}
\end{equation}
where $I=A^2$ is the classical intensity. The frequency shift depends explicitly on both the dimensionless GUP parameter $\beta_{\rm NL}$ and the dimensionless oscillation amplitude $A$. Consequently, a quantitative determination of $\beta_{\rm NL}$ can be achieved through precise measurements of the oscillation amplitude and the resulting frequency shift.

It is important to emphasize that the auxiliary momentum $\hat{p}$ and the associated nonlinear Hamiltonian~\eqref{eq:non_Hamilton_b} are purely mathematical constructs introduced for formal convenience. Physically, the momentum and Hamiltonian of the oscillator remain those of the standard harmonic oscillator, expressed in terms of the physical operator $\hat{\tilde{p}}$ and the corresponding quadratic form in Eq.~\eqref{eq:Hamilton_b}.  Accordingly, the auxiliary momentum $\hat{p}$ is not a directly measurable observable and carries no additional covariance requirements. As will be discussed in detail in the following sections, the entire measurement scheme relies exclusively on the detection of the displacement quadrature $\hat{q}$.

\section{The model and the Langevin equation}
\label{The model and the Langevin equation}
\begin{figure}[]
\centering
\includegraphics[width=3.2in]{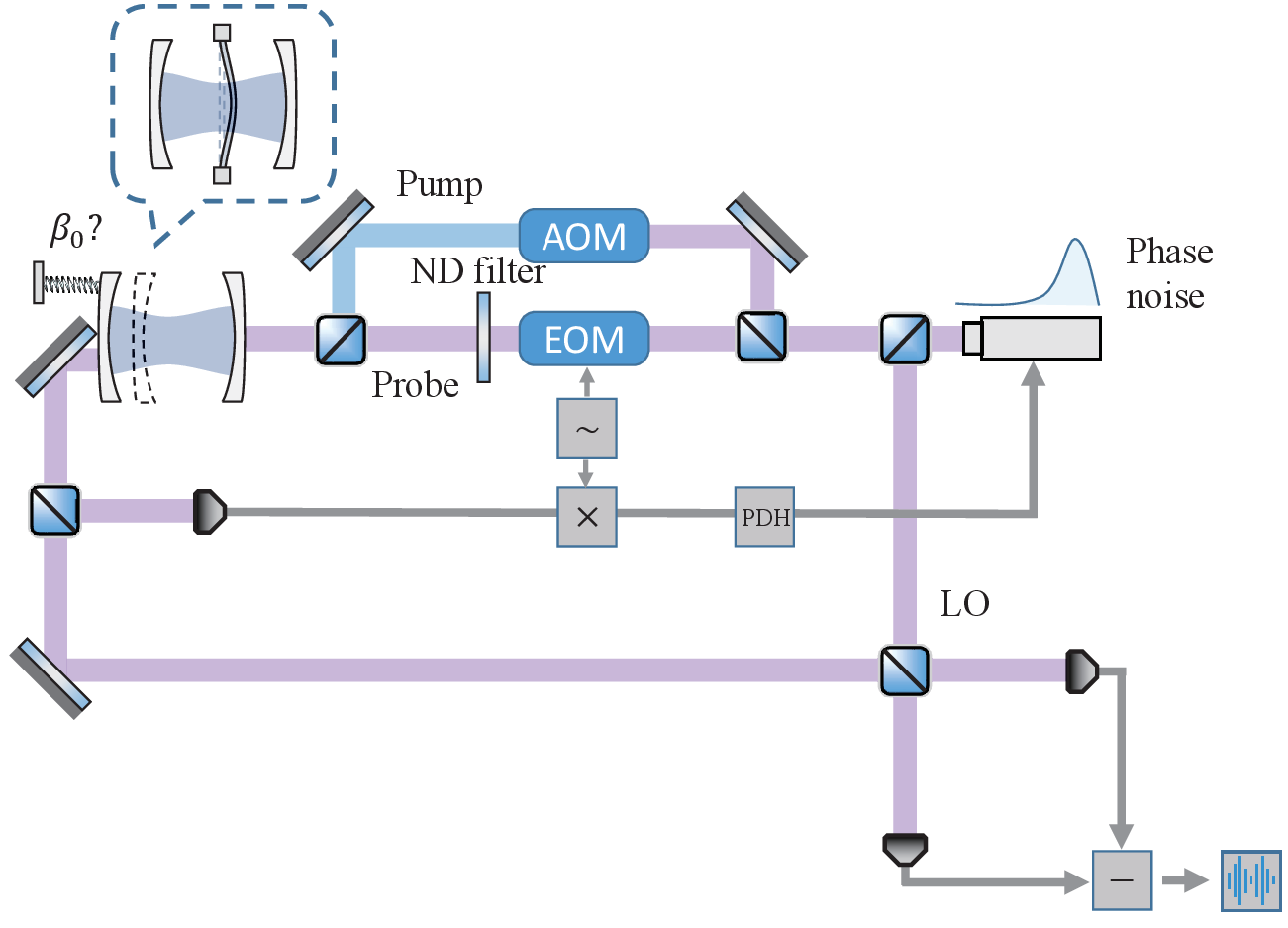}  
\caption{Schematic of the optomechanical detector. A probe beam is incident on an optomechanical cavity that can be configured either as a FP cavity terminated by a movable end mirror or as a MiM system with a thin dielectric membrane placed between two fixed mirrors. The probe beam is phase-modulated by an EOM and locked to cavity resonance using the PDH technique. The reflected probe light is analyzed by balanced homodyne detection to monitor the mechanical displacement and extract the nonlinear parameter $  \beta_{\rm NL}  $. A second pump beam is simultaneously injected into the cavity to drive and manipulate the mechanical oscillator.
\label{fig:1}}
\end{figure}

As illustrated in Fig.~\ref{fig:1}, we consider a measurement optical configuration that described in Refs.~\cite{Bonaldi2020,Li2025}, in which a laser beam passes through an electro-optic modulator (EOM) to generate frequency-detuned components. These components are injected into one port of the OMS cavity as the driving field (mode $1$) and the probe field (mode $2$), respectively. The driving field is modulated to excite the mechanical resonator, while the probe field is phase-modulated to generate a calibration signal used to quantify the resonator dynamics. We examine two types of cavity OMS, as depicted in the figure: (1) a cavity in which one mirror is fixed and the other acts as a movable mirror~\cite{Caves1980,Corbitt2006}, and (2) a membrane-in-the-middle configuration in which a thin membrane is placed inside the cavity~\cite{Jayich2008,Xuereb2011,Lij2016,Piergentili2018,Sheng2020,Cao2025}. At the level of first-order optomechanical coupling, these two systems exhibit equivalent dispersive coupling forms; however, they differ in the physical mechanisms responsible for generating higher-order couplings~\cite{Lij2016,Li20262}. The transmitted cavity field (which possesses identical properties to the reflected field, without loss of generality) serves as the output field and is subjected to homodyne detection~\cite{Bonaldi2020,Li2025}. The Hamiltonian of the entire system can then be expressed as: $H=H_a+H_{m}+H_{\rm pump}+H_{\rm probe}$, where 
\begin{equation}
\begin{split}
H_a/\hbar=\left[\omega_c\left(X\right)\right]\left(\hat{a}_1^\dagger \hat{a}_1+\hat{a}_2^\dagger \hat{a}_2\right),
\end{split}
\label{eq:free_Hamilton}
\end{equation}
is the free Hamiltonian of the cavity field. Here, $  \omega_c(X)  $ denotes the position-dependent resonance frequency of the cavity modes. The operators $\hat{a}_j$ ($\hat{a}_j^\dagger$) ($j=1,2$) are the corresponding annihilation (creation) operators for the cavity field. The free Hamiltonian of the mechanical oscillator, expressed in the auxiliary-momentum representation, incorporates the GUP-induced nonlinear correction terms and reads
\begin{equation}
\begin{split}
H_{m}/\hbar=\dfrac{\omega_b}{2}\left(\hat{q}^2+\hat{p}^2\right)+\dfrac{1}{3}\omega_b\beta_{\rm NL}\hat{p}^4.
\end{split}
\label{eq:free_Hamilton_b}
\end{equation}
Correspondingly, the driving and probing Hamiltonians are given by
\begin{equation}
\begin{split}
H_{\rm pump}/\hbar=&i E_1(\hat{a}_1^\dagger e^{-i\omega_{d_1} t}-\hat{a}_1e^{i\omega_{d_1} t})\\&+i E_2(\hat{a}_1^\dagger e^{-i\omega_{d_2} t}-\hat{a}_1e^{i\omega_{d_2} t}),
\end{split}
\label{eq:Hamilton_pump}
\end{equation}
and 
\begin{equation}
\begin{split}
H_{\rm probe}/\hbar=&i E_p(\hat{a}^\dagger e^{-i\omega_pt-i\phi\sin\Omega t}-\hat{a}e^{i\omega_pt+i\phi\sin\Omega t}),
\end{split}
\label{eq:Hamilton_probe}
\end{equation}
respectively. Here, $E_o$ ($o=1,2,p$) denotes the driving amplitude of the corresponding optical field and satisfies $E_o=\sqrt{2\kappa_{\rm in}P_o/\hbar\omega_o}$, where $P_o$ is the input laser power and $\kappa_{\rm in}$ is the cavity decay rate through the input port.

Following the procedure outlined in Ref.~\cite{Li2025}, the driving fields are frequency-modulated at $\omega_{d_1}$ and $  \omega_{d_2}$ to excite the mechanical resonator~\cite{Li2025,Li2026,Li20262,Mari2009}, while the probe field at frequency $\omega_p$ is phase-modulated with depth $\phi$ at frequency $\Omega$ to generate a calibration signal for quantifying the resonator dynamics~\cite{Gorodetsky2010}. Lasers inevitably exhibit phase noise $\Phi$, which is predominantly concentrated in the phase quadrature and characterized by a finite linewidth $\Gamma_l$. Amplitude noise, by contrast, is negligible. A more realistic description of this noise is provided by the following bandpass filtered frequency noise spectrum~\cite{Abdi2011}:
\begin{equation}
\begin{split}
\mathcal{S}_\Phi(\omega_p)=2\Gamma_l\dfrac{\Omega_p^4}{(\Omega_p^2-\omega)^2+\omega^2\tilde{\gamma}^2},
\end{split}
\label{eq:autocorrelation function}
\end{equation}
where $\Omega$ denotes the center frequency of the noise band, $\tilde{\gamma}$ represents the bandwidth of the frequency-noise spectrum, and $\Gamma_l$ quantifies the noise intensity that determines the laser linewidth. Following an approach analogous to the pseudomode formalism, the colored phase noise $\Phi$ is represented by white noise through the introduction of an auxiliary mode. Using the full and exact Hamiltonians given above together with the fluctuation–dissipation processes for both the cavity and mechanical modes, we obtain the following Heisenberg-Langevin equations~\cite{Li20252,Aspelmeyer2014,noise}:
\begin{equation}
\begin{split}
\dot{\hat{a}}_1=&\left[-\kappa-i\omega_c\left(X\right)+i{\psi}_1\right]\hat{a}_1\\&+E_1e^{-i\omega_{d_1}t}+E_2e^{-i\omega_{d_2}t}+\sqrt{2\kappa}\hat{a}_1^{in},\\
\dot{\hat{a}}_2=&\left[-\kappa-i\omega_c\left(X\right)+i{\psi}_2\right]\hat{a}_2\\&+E_pe^{-i\omega_pt-i\phi\sin\Omega t}+\sqrt{2\kappa}\hat{a}_2^{in},\\
\dot{\hat{q}}=&\omega_b \hat{p}\left(1+\dfrac{4}{3}\beta_{\rm NL}\hat{p}^2\right),\\
\dot{\hat{p}}=&-\omega_b \hat{q}-\gamma \hat{p}\\&+\left[-\dfrac{\partial [\omega_c(X)]}{\partial q}\right]\left(\hat{a}_1^\dagger \hat{a}_1+\hat{a}_2^\dagger \hat{a}_2\right)+\hat{\xi},\\
\dot{\psi}_j=&\Omega_p\theta_j,\\
\dot{\theta}_j=&-\Omega_p\psi_j-\tilde{\gamma}\theta_j+\Omega\sqrt{2\Gamma_l}\xi_{p,j},
\end{split}
\label{eq:QLE}
\end{equation}
where $\kappa=\kappa_{\rm in}+\kappa_{\rm ex}$ is the total decay rate of cavity mode, with $\kappa_{\rm ex}$ the optical loss rate through all ports different from the input port, and $\gamma$ the mechanical decay rate. $\psi:=\dot{\Phi}$ is the phase noise defined by Eq.~\eqref{eq:autocorrelation function}, and $\theta$ is the introduced auxiliary mode.The input noise operators $\hat{a}_j^{\rm in}$ ($  j=1,2$) and $\hat{\xi}$ are zero-mean reservoir noise operators satisfying the following white Gaussian correlation functions~\cite{Giovannetti2001}
\begin{equation}
\begin{split}
&\langle \{\hat{a}_j^{in}(t)^\dagger,\hat{a}_{j'}^{in}(t')\}\rangle= (2\bar{n}_a+1)\delta_{jj'}\delta(t-t'),\\
&\langle \hat{\{\xi}(t),\hat{\xi}'(t')\}\rangle=2\gamma(2\bar{n}_{b}+1)\delta(t-t').
\end{split}
\label{eq:noise autocorrelation function}
\end{equation}
All other second-order correlation functions vanish. Correspondingly, $\xi_p$ is a zero-mean Gaussian white-noise process satisfying the correlation function $\langle\xi_{p,j}(t)\xi_{p,j}(t')\rangle=\delta_{jj'}\delta(t-t')$. In the weakly nonlinear regime, the Gaussian dynamics of the system can be described by transforming the quantum Langevin equations into $c$-number stochastic Langevin equations:
\begin{equation}
\begin{split}
\dot{a}_1=&\left[-\kappa-i\omega_c\left(X\right)+i{\psi_1}\right]{a}_1\\&+E_1e^{-i\omega_{d_1}t}+E_2e^{-i\omega_{d_2}t}+\sqrt{2\kappa}{a}_1^{in},\\
\dot{a}_2=&\left[-\kappa-i\omega_c\left(X\right)+i{\psi_2}\right]{a}_2\\&+E_pe^{-i\omega_pt-i\phi\sin\Omega t}+\sqrt{2\kappa}{a}_2^{in},\\
\dot{{q}}=&\omega_b {p}\left(1+\dfrac{4}{3}\beta_{\rm NL}{p}^2\right),\\
\dot{{p}}=&-\omega_b {q}-\gamma {p}\\&+\left[-\dfrac{\partial [\omega_c(X)]}{\partial q}\right]\left(\vert {a}_1\vert^2+\vert {a}_2\vert^2-1\right)+{\xi},\\
\dot{\psi}_j=&\Omega_p\theta_j,\\
\dot{\theta}_j=&-\Omega_p\psi_j-\tilde{\gamma}\theta_j+\Omega\sqrt{2\Gamma_l}\xi_{p,j},
\end{split}
\label{eq:CLE}
\end{equation}
Here, a correction term of $-1$ has been added to the equation for $\dot{p}$. This arises because the $c$-number variables employed do not obey the canonical commutation relations of the original operators~\cite{Rodrigues2010}. Accordingly, the correlation functions of the noise operators have been modified simultaneously to~\cite{Wang2014,Lee2013,Li2020,Li2017}
\begin{equation}
\begin{split}
&\langle {a}_j^{in}(t)^*{a}_{j'}^{in}(t')\rangle= \left(\bar{n}_a+\dfrac{1}{2}\right)\delta_{jj'}\delta(t-t'),\\
&\langle \xi(t)\xi'(t')\rangle=\gamma(2\bar{n}_{b}+1)\delta(t-t').
\end{split}
\label{eq:c noise autocorrelation function}
\end{equation}
This approximation preserves all classical nonlinear properties while neglecting nonlinear quantum effects that can render the Wigner function negative~\cite{Lee2013}. Consequently, Eq.~\eqref{eq:CLE} is valid for a driven-dissipative optomechanical system provided that the system parameters allow the dynamics to be approximated as linear, or when the driving is sufficiently strong. In this regime, the Wigner function of the mechanical oscillator remains non-negative. In contrast, when the oscillator energy comprises only a few quanta, the nonlinearity becomes significant and can drive the Wigner function negative, thereby rendering Eq.~\eqref{eq:CLE} invalid.

\subsection{Fabry-P\'erot optomechanics}
\label{Fabry-Perot OMS}
For a FP OMS, displacement of the movable mirror modifies the effective cavity length, thereby shifting the cavity resonance frequency. Consequently, the position-dependent frequency $\omega_c(X)$ takes the form~\cite{Aspelmeyer2014,Gao2015}:
\begin{equation}
\begin{split}
\omega_c(X)=\dfrac{sc\pi}{L+X}=\dfrac{sc\pi}{L+x_{\rm ZPF} q},
\end{split}
\label{eq:FP_omegax}
\end{equation}
Here, $L$ is the length of the cavity and $m$ is an integer factor. Substituting the above expression for $\omega_c(X)$ into the stochastic Langevin equations~\eqref{eq:CLE} yields the complete set of dynamical equations for the Fabry-P\'erot OMS, including all relevant noise terms:
\begin{equation}
\begin{split}
\dot{a}_1=&\left[-\kappa-i\left(\dfrac{sc\pi}{L+x_{\rm ZPF} q}-\omega_{d_1}+{\psi}_1\right)\right]{a}_1\\&+E_1+E_2e^{-i(\omega_{d_2}-\omega_{d_1})t}+\sqrt{2\kappa}{a}_1^{in},\\
\dot{a}_2=&\left[-\kappa-i\left(\dfrac{sc\pi}{L+x_{\rm ZPF} q}-\omega_{p}-{\phi\Omega}\cos\Omega t+{\psi}_2\right)\right]{a}_2\\&+E_p+\sqrt{2\kappa}{a}_2^{in},\\
\dot{{q}}=&\omega_b {p}\left(1+\dfrac{4}{3}\beta_{\rm NL}{p}^2\right),\\
\dot{{p}}=&-\omega_b {q}-\gamma {p}\\&+\left[\dfrac{sc\pi x_{\rm ZPF}}{\left(L_0+x_{\rm ZPF} q\right)^2}\right]\left(\vert {a}_1\vert^2+\vert {a}_2\vert^2-1\right)+{\xi},\\
\dot{\psi}_j=&\Omega_p\theta_j,\\
\dot{\theta}_j=&-\Omega_p\psi_j-\tilde{\gamma}\theta_j+\Omega\sqrt{2\Gamma_l}\xi_{p,j}.
\end{split}
\label{eq:FP_CLE}
\end{equation}
Here, we have rotated the frames of cavity modes $a_1$ and $a_2$ with respect to $e^{-i\omega_{d_1}t}$ and $e^{-i\omega_pt-i\phi\sin\Omega t}$, respectively. Expanding Eq.~\eqref{eq:FP_omegax} yields:
\begin{equation}
\begin{split}
\omega_c(X)=\omega^{\rm F}_0+\sum_{n=1}^{\infty}q^n\dfrac{1}{n!}\left.\dfrac{\partial^n \omega_c}{\partial q^n}\right|_{q=0},
\end{split}
\label{eq:F_expansion}
\end{equation}
where $\omega_0^{\rm F}=n\pi c/L$ is the cavity resonance frequency. The higher-order optomechanical coupling coefficients are defined as
\begin{equation}
\begin{split}
g^{\rm F}_{n-1}:=-\dfrac{1}{2^{n/2}n!}\left.\dfrac{\partial^n \omega_c}{\partial q^n}\right|_{q=0},
\end{split}
\label{eq:FP_g}
\end{equation}
with the leading coefficient
\begin{equation}
\begin{split}
g^{\rm F}_{0}=\dfrac{x_{\rm ZPF}}{\sqrt{2}}\dfrac{sc\pi}{L^2},
\end{split}
\label{eq:FP_g0}
\end{equation}
recovering the single-photon optomechanical coupling strength of the standard linearized OMS model.
\subsection{Membrane-in-middle optomechanics}
\label{Membrane Optomechanics}
For the MiM OMS, the membrane positioned inside the Fabry–P\'erot cavity perturbs the intracavity field distribution, thereby inducing a position-dependent frequency shift $\delta\omega$ of the cavity resonance. The length, width, and thickness of the membrane are denoted by $L_x$, $L_y$, and $L_z$, respectively; its density is $\rho$ and its refractive index is $n$. The position-dependent cavity frequency is therefore given by
\begin{equation}
\begin{split}
\omega_c(X)=\dfrac{s c \pi}{L}+\delta\omega,
\end{split}
\label{eq:M_omegax}
\end{equation}
where $\omega_0=sc\pi/L$ is still the unperturbed cavity resonance frequency. This shift $\delta\omega$ originates from the spatial redistribution of the electromagnetic field inside the cavity, a subject that has been analyzed in detail in Refs.~\cite{Jayich2008,Lij2016,Piergentili2018,Li20252}. Here, we write $X$ as $X=Q+x_{\rm ZPF}q$, where $Q$ serves to characterize the equilibrium position where the thin film was initially placed. Following their derivation, we adopt the explicit expression
\begin{equation}
\begin{split}
&\delta\omega(Q,q)=\dfrac{c}{L}\left\{(-1)^l\arcsin\left[\mathcal{F}\left(Q,q\right)\right]\right\},
\end{split}
\label{eq:delta_omega}
\end{equation}
with 
\begin{equation}
\begin{split}
\mathcal{F}\left(Q,q\right)=\sqrt{R}\cos\left[2k\left(Q+x_{\rm ZPF}q\right)\right].
\end{split}
\label{eq:F}
\end{equation}
Here $k=2\pi/\lambda$ is the wave number of the electric field, and $\lambda$ is its wavelength.  The complex amplitude reflection coefficient $r$ of the membrane can be expressed in terms of its refractive index $n$ and thickness $L_z$ as
\begin{equation}
\begin{split}
\sqrt{R}e^{i\phi}=r=\dfrac{(n^2-1)\sin(knL_z)}{(n^2+1)\sin(knL_z)+2in\cos(knL_z)},
\end{split}
\label{eq:R}
\end{equation}
Here we have omitted the corresponding expression of transmittance, which is also provided  in Ref.~\cite{Piergentili2018}, because the subsequent discussion does not require a calculation based on transmittance.  Substituting Eqs.~\eqref{eq:M_omegax}-\eqref{eq:F} into Eq.~\eqref{eq:CLE}, and employing the following relationship,
\begin{widetext}
\begin{equation}
\begin{split}
-\dfrac{\partial [\delta\omega]}{\partial q}=-\dfrac{\partial [\delta\omega]}{\partial X}\dfrac{\partial X}{\partial  q_1}=x_{\rm ZPF}\dfrac{c}{L}\left[(-1)^{l+1}\dfrac{2k\sqrt{R}\sin\left[2k\left(Q+x_{\rm ZPF}q\right)\right]}{\sqrt{1-R\cos^2\left[2k\left(Q+x_{\rm ZPF}q\right)\right]}}\right]:=\mathcal{L}(Q,q),
\end{split}
\label{eq:L1}
\end{equation}
we obtain:
\begin{equation}
\begin{split}
\dot{a}_1=&\left[-\kappa-i\left(\dfrac{s c \pi}{L_0}+\delta\omega(Q,q)-\omega_{d_1}+{\psi}_1\right)\right]{a}_1+E_1+E_2e^{-i(\omega_{d_2}-\omega_{d_1})t}+\sqrt{2\kappa}{a}_1^{in},\\
\dot{a}_2=&\left[-\kappa-i\left(\dfrac{s c \pi}{L_0}+\delta\omega(Q,q)-\omega_{p}-{\phi\Omega}\cos\Omega t+{\psi}_2\right)\right]{a}_2+E_p+\sqrt{2\kappa}{a}_2^{in},\\
\dot{{q}}=&\omega_b {p}\left(1+\dfrac{4}{3}\beta_{\rm NL}{p}^2\right),\\
\dot{{p}}=&-\omega_b {q}-\gamma {p}+\mathcal{L}(Q,q)\left(\vert {a}_1\vert^2+\vert {a}_2\vert^2-1\right)+{\xi},\\
\dot{\psi}_j=&\Omega_p\theta_j,\\
\dot{\theta}_j=&-\Omega_p\psi_j-\tilde{\gamma}\theta_j+\Omega\sqrt{2\Gamma_l}\xi_{p,j}.
\end{split}
\label{eq:MB_CLE}
\end{equation}
\end{widetext}
Consistent with the case of the Fabry–Pérot cavity OMS, we also have rotated the frames of cavity modes $a_1$ and $a_2$ with respect to $e^{-i\omega_{d_1}t}$ and $e^{-i\omega_pt-i\phi\sin\Omega t}$, respectively. We also expanding Eq.~\eqref{eq:M_omegax} as:
\begin{equation}
\begin{split}
&\delta\omega(Q,q)=\delta\omega(Q,0)+\sum_{n=1}^{\infty}q^n\dfrac{1}{n!} \left.\dfrac{\partial^n [\delta\omega]}{\partial q^n}\right|_{Q,0}.
\end{split}
\label{eq:M_expansion}
\end{equation}
The renormalized cavity resonance frequency is then
\begin{equation}
\begin{split}
\omega^{M}_0=\dfrac{s c \pi}{L_0}+\delta\omega(Q,0),
\end{split}
\label{eq:M_omega0}
\end{equation}
while the higher-order optomechanical coupling coefficients are defined by
\begin{equation}
\begin{split}
g^{\rm M}_{n-1}:=\dfrac{1}{2^{n/2}n!}\left.\dfrac{\partial^n [\delta\omega]}{\partial q^n}\right|_{Q,0},
\end{split}
\label{eq:M_g}
\end{equation}
and the single-photon OMS coupling strength is
\begin{equation}
\begin{split}
g^{\rm M}_0=\dfrac{1}{\sqrt{2}}\mathcal{L}(Q,0).
\end{split}
\label{eq:M_g0}
\end{equation}
Eqs.~\eqref{eq:FP_CLE} and~\eqref{eq:MB_CLE} are the basis for our subsequent calculations.
\section{Quantitative measurement of GUP}
\label{Measuring GUP in the quantum regime}
The measurement protocol proceeds in two stages~\cite{Li2025}. At the initial time $t=0$, the probe, cooling, and excitation fields are simultaneously activated. After a preparation interval of duration $t_p$, the mechanical oscillator reaches a highly pure and strongly excited state. The cooling and excitation fields are then deactivated, and the output spectrum of the probe field is acquired over the subsequent measurement window $t\in(t_p, t_p + \Delta t]$. The temporal profile of the driving-field intensities is given by
\begin{equation}
\left\{
\begin{split}
    &E_{1,2}(t)=E_{1,2},\,E_p(t)=E_p,\,\,(t\in[0,t_p])\\
    &E_{1,2}(t)=0,\,E_p(t)=E_p,\,\,(t\in[t_p, t_p + \Delta t])
\end{split}
\right.
\label{eq:power_time}
\end{equation}
The corresponding frequency detuning is given by:
\begin{equation}
\begin{split}
&\Delta_1=\omega_0+x_{\rm PDH}-\omega_{d_1},\\
&\Delta_2=\omega_{d_2}-\omega_{d_1},\\
&\Delta_p=\omega_0+x_{\rm PDH}-\omega_{d_p},
\end{split}
\label{eq:detunting}
\end{equation}
Here $  x_{\rm PDH}  $ is the frequency correction imposed by the Pound–Drever–Hall (PDH) cavity-locking servo~\cite{Bonaldi2020,Piergentili2021}. In practice, active PDH feedback is required to compensate laser-frequency drift and to maintain the laser at the prescribed sideband condition. Because the servo cannot distinguish the bare cavity resonance from the low-frequency component of the cavity-frequency shift induced by the oscillator motion, this correction, to first order in the low-frequency part of the oscillator dynamics, takes the explicit form~\cite{Li2025}
\begin{equation}
\begin{split}
x_{\rm PDH}(t)=\dfrac{1}{\tau}\int_t^{t+\tau}g_0q(t)dt,
\end{split}
\label{eq:x_pdh}
\end{equation}
where $q(t)$ is the oscillator displacement, $g_0$ is the first order optomechanical coupling rate in Eqs.~\eqref{eq:FP_g} and~\eqref{eq:M_g}, and $\tau$ is the effective averaging time set by the locking bandwidth.

The output field transmitted from the cavity port opposite the input laser obeys the input-output boundary condition:
\begin{equation}
\begin{split}
a_{\rm out}=\sqrt{2\kappa_{\rm in}}a_2-a^{in}_2,
\end{split}
\label{eq:input-output}
\end{equation}
The phase quadrature of this output field is extracted by balanced homodyne detection, yielding the photocurrent signal $\alpha_{\rm out}=K{\rm Im}(a_{\rm out})$, with $K$ the overall detection gain. The spectrum of the corresponding output field is expressed as:
\begin{equation}
\begin{split}
S_{\rm out}(\omega)=\dfrac{1}{2\pi}\int d\tau\langle \alpha_{\rm out}(t)\alpha_{\rm out}^*(0) \rangle e^{-i\omega t},
\end{split}
\label{eq:spectrum}
\end{equation}
In the experiment, this spectrum is evaluated from the Fourier transform of the measured time-domain photocurrent signal according to
\begin{equation}
\begin{split}
S_{\rm out}'(\omega)=\dfrac{1}{2\pi}\left\vert \int d\tau\langle \alpha_{\rm out}(t)\rangle e^{-i\omega\tau}\right\vert^2.
\end{split}
\label{eq:FFT_spectrum}
\end{equation}
The measured output spectrum exhibits a characteristic bimodal structure. A calibration peak appears at the frequency-modulation frequency $\Omega$ of the probe field; the height of this peak is recorded and used to calibrate the displacement amplitude of the oscillator. Near the resonance frequency $\omega_b$ lies the signal peak. The height of the signal peak is likewise recorded for amplitude calibration, while its precise center frequency encodes the GUP correction. By varying the driving field intensity across a series of measurements, the oscillator is excited to different steady-state amplitudes satisfying $A\propto E_1E_2$~\cite{Li2026,Mari2009}. For each amplitude, the corresponding oscillation frequency is extracted from the position of the signal peak. With the amplitude independently calibrated, the measured peak frequency is plotted as a function of the intensity $I$ and fitted to the linear relation predicted by Eq.~(\ref{eq:eff})~\cite{Li2025,Li20262}. The slope of this fit directly determines the dimensionless GUP parameter $  \beta_{\rm NL}  $.

\begin{table}[]
\caption{Parameter}
        \centering
       \begin{tabular}{cccc}
         \hline
       Parameter & Value & Dimensionless\\ 
          \hline
         $\omega_b/2\pi$  & $525$\,kHz & $1$&  \\
         $\Delta_1/2\pi$  & $675$\,kHz & $ 1.2857$  \\
         $\Delta_2/2\pi$  & $522$\,kHz & $0.9943$  \\
         $\gamma/2\pi$    &  $5.25\times 10^{-4}$\,Hz & $ 10^{-9}$ \\
         $\kappa_{\text{in}}/2\pi$  & $50$\,kHz & $0.0952$ \\
         $\kappa_{\text{ex}}/2\pi$  & $50$\,kHz & $0.0952$ \\
         $\lambda$& $1064$\,nm & \\  
         $n$& $2.17$ &  \\
         $m^F$& $5$\,ng  &\\
         $L^{\rm M}$& $0.03$\,m & \\        
         $L^{\rm M}_{x,y,z}$& $0.67$\,mm,$0.67$\,mm,$142$\,nm &\\
         $\rho^M$& $3100$\,kg\,m$^{-3}$  &\\
         \hline
       \end{tabular}
       \label{tab}
\end{table}

The parameters of the OMSs considered in this work coincide with the values reported in the Ref.~\cite{Li2025} under the first order approximation, as summarized in Tab.~\ref{tab}. This choice eliminates the effects of extraneous parameter variations and thereby enables a direct comparison with the results of that reference. With the exception of the oscillator quality factor, which we take to be higher than the values realized in current GUP experiments, the parameters are otherwise consistent with those employed in existing experimental demonstrations of GUP measurements.

\subsection{Probe-field phase noise in the output spectrum} 
\begin{figure}[]
\centering
\includegraphics[width=3in]{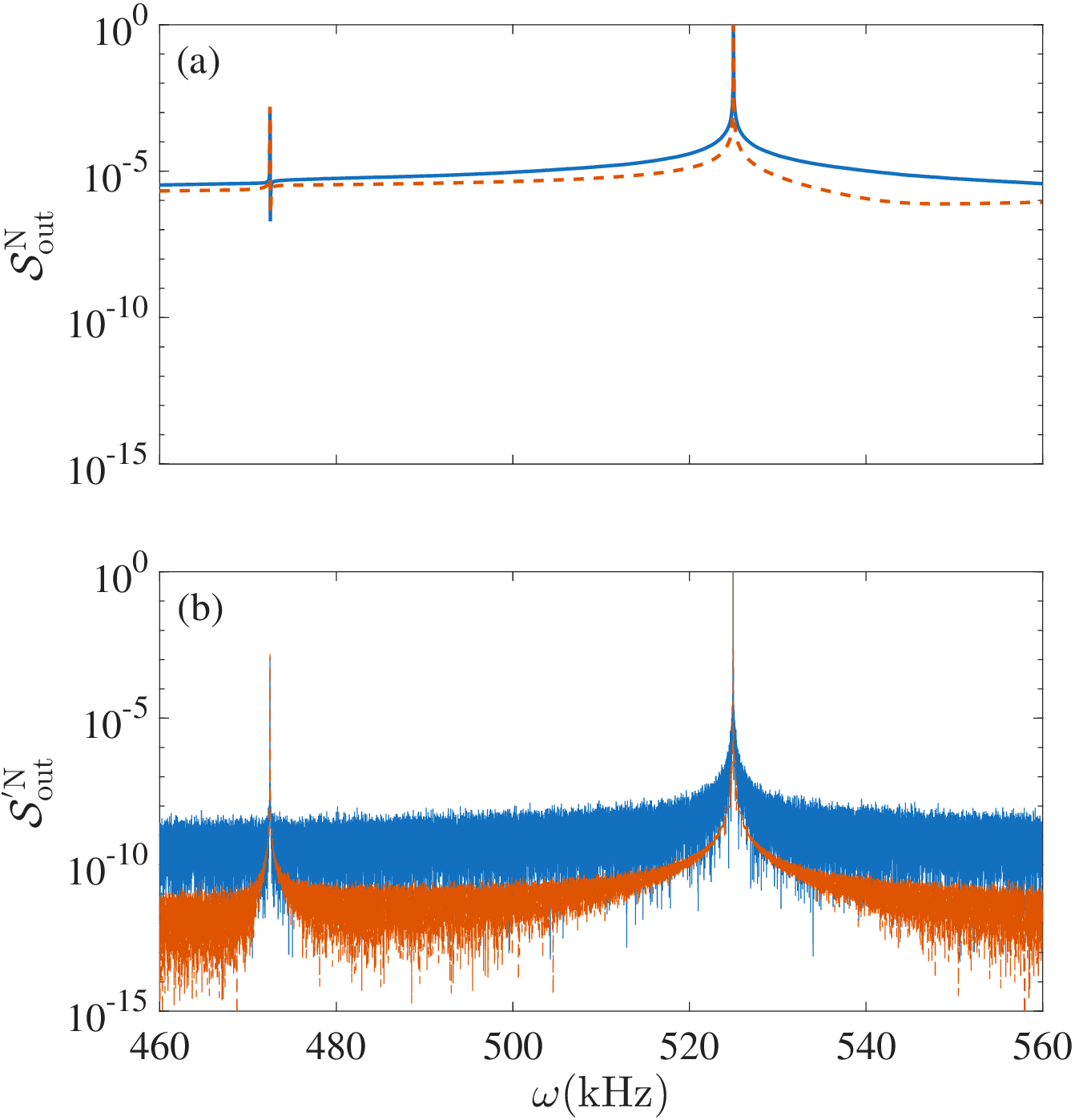}  
\caption{Comparison of the output spectra of the probe field with and without phase noise. (a): Power spectrum obtained from Eq.~\eqref{eq:spectrum}. (b): Spectrum calculated via the FFT according to Eq.~\eqref{eq:FFT_spectrum}. Each spectrum is normalized to its maximum value, that is, $\mathcal{S}^{\rm N}_{\rm out}=\mathcal{S}_{\rm out}/\max\{\mathcal{S}_{\rm out}\}$ and $\mathcal{S'}^{\rm N}_{\rm out}=\mathcal{S'}_{\rm out}/\max\{\mathcal{S}_{\rm out}\}$. In addition to the parameters listed in Table~\ref{tab}, we set $\beta_{\rm NL}=0$, $\Omega/\omega_b=0.9$, $\phi=0.01$, $\gamma\tau=5.25\times 10^{-6}$, $P_1=2\,\mu\mathrm{W}$, $P_2=50\,\mu\mathrm{W}$, and $P_p=5\,\mu\mathrm{W}$.
\label{fig:2}}
\end{figure}
From Eqs.~\eqref{eq:FP_CLE} and~\eqref{eq:MB_CLE}, the presence of phase noise under the first-order approximation modifies the probe-field equation by adding a white-noise term $\psi_2$ to the detected signal. Consequently, the signal transforms as:
\begin{equation}
\begin{split}
g_0q + \phi\Omega\cos\Omega t\rightarrow  g_0q + \phi\Omega\cos\Omega t-\psi_2
\end{split}
\label{eq:signal transforms}
\end{equation}
This additional noise component could, in principle, obscure the oscillator signal peak. Such an effect can arise even in measurement schemes that impose no extra requirements, for example the original proposal of Bawaj et al. for detecting a classical oscillator. However, as shown in Fig.~\ref{fig:2}, for the parameters employed in this work the phase noise remains sufficiently weak that it does not mask the oscillator peak. Therefore, the influence of phase noise can be neglected in the present classical measurement scheme.

\subsection{Quantitative impact of high-order nonlinearities} 
We next examine the influence of higher-order optomechanical interactions on the measurement scheme. Substituting Eqs.~\eqref{eq:F_expansion} and~\eqref{eq:M_expansion} into the equation of motion for the mechanical oscillator and introducing, for convenience, we define the complex variable $b:=(q+ip)/\sqrt{2}$ which satisfies $\vert b\vert=A$. We then obtain, after neglecting fluctuation-dissipation processes, the following dynamical equation:
\begin{equation}
\begin{split}
\dot{b}\simeq & -i\omega_bb+i\omega_b\dfrac{\beta_{\rm NL}}{3}\left(b-b^*\right)^3\\&+i\vert {a}_2\vert^2\left[\sum^{\infty}_{n=1}n!g_{n-1}\left(b+b^*\right)^{n-1}\right].\\
\end{split}
\label{eq:P_CLE}
\end{equation}
Under the conditions that the oscillator undergoes single-mode vibration near its intrinsic frequency and that rapidly oscillating terms are neglected within the rotating-wave approximation, the dynamical equation simplifies to
\begin{equation}
\begin{split}
\dot{b}\simeq & -i\left[\omega_b+\left(\beta_{\rm NL}-12g_3\vert a_2\vert^2\right)\vert b\vert^2-2g_1\vert {a}_2\vert^2\right]b\\&+i\vert {a}_2\vert^2\left(g_{0}+6g_2\vert b\vert^2\right)+\mathcal{O}(g_4).\\
\end{split}
\label{eq:P_CLE}
\end{equation}
under the condition of the oscillator is in single-mode vibration at a frequency close to its intrinsic frequency, and neglecting higher-order rotating-wave terms. Under the conditions of zero detuning of the probe field and single-mode oscillator dynamics, the effective frequency of the oscillator is
\begin{equation}
\begin{split}
\omega_{\rm eff}=\omega_b+\left(\beta_{\rm NL}-12g_3\vert a_2\vert^2\right)I-2g_1\vert {a}_2\vert^2,
\end{split}
\label{eq:omega_eff_full}
\end{equation}
For zero probe detuning and single-mode oscillator dynamics, the effective frequency of the oscillator is therefore
\begin{equation}
\begin{split}
\omega_{\rm eff}=\omega_b+\beta_{\rm NL}I,
\end{split}
\label{eq:omega_eff}
\end{equation}
which coincides with the result obtained for an isolated mechanical oscillator in the absence of the probe field. Comparing Eqs.~\eqref{eq:omega_eff_full} and \eqref{eq:omega_eff}, we find that the parameters $\beta_{\rm NL}$ and $g_3\vert a_2\vert^2$ combine into a effective coefficient. As a result, the standard measurement scheme based on the amplitude dependence of the frequency shift cannot distinguish between the two contributions. It is also important to note that the term $-2g_1|a_2|^2$ is independent of the mechanical amplitude. This term therefore contributes only a constant offset to $\omega_{\rm eff}$ and does not affect the slope obtained from the linear fit, provided that the intracavity photon number of the probe field remains constant while the drive power is varied. In practice, this condition must be carefully ensured. If the probe and drive fields originate from the same laser and are split into two paths, any adjustment of the drive power will generally change the probe power simultaneously unless additional stabilization or independent control is implemented. Because $E_1$, $E_2$, and $E_p$ are derived from the same source, they satisfy $E_1\propto E_2\propto E_p$. Combined with the relations $  A\propto E_1E_2  $ and $\vert a_2\vert \propto E_p^2$, the full expression for the effective frequency can be rewritten solely in terms of the mechanical amplitude as
\begin{equation}
\begin{split}
\omega_{\rm eff}=\omega_b+\left(\beta_{\rm NL}-12g_3\vert a_2\vert^2\right)I-2Kg_1\sqrt{I},
\end{split}
\label{eq:omega_eff_full_a}
\end{equation}
where $K$ is a proportionality constant that depends on the power intensity, cavity quality factor and the OMS coupling strengths. In the regime $Kg_1/\omega_b\ll 1$, the additional term $-2Kg_1\sqrt{I}$ introduces a correction to the slope of the linear fit of $\omega_{\rm eff}$ versus $  I$. To leading order, this correction modifies the extracted slope by a relative factor of order $Kg_1/2$.

\begin{figure}[]
\centering
\includegraphics[width=3.4in]{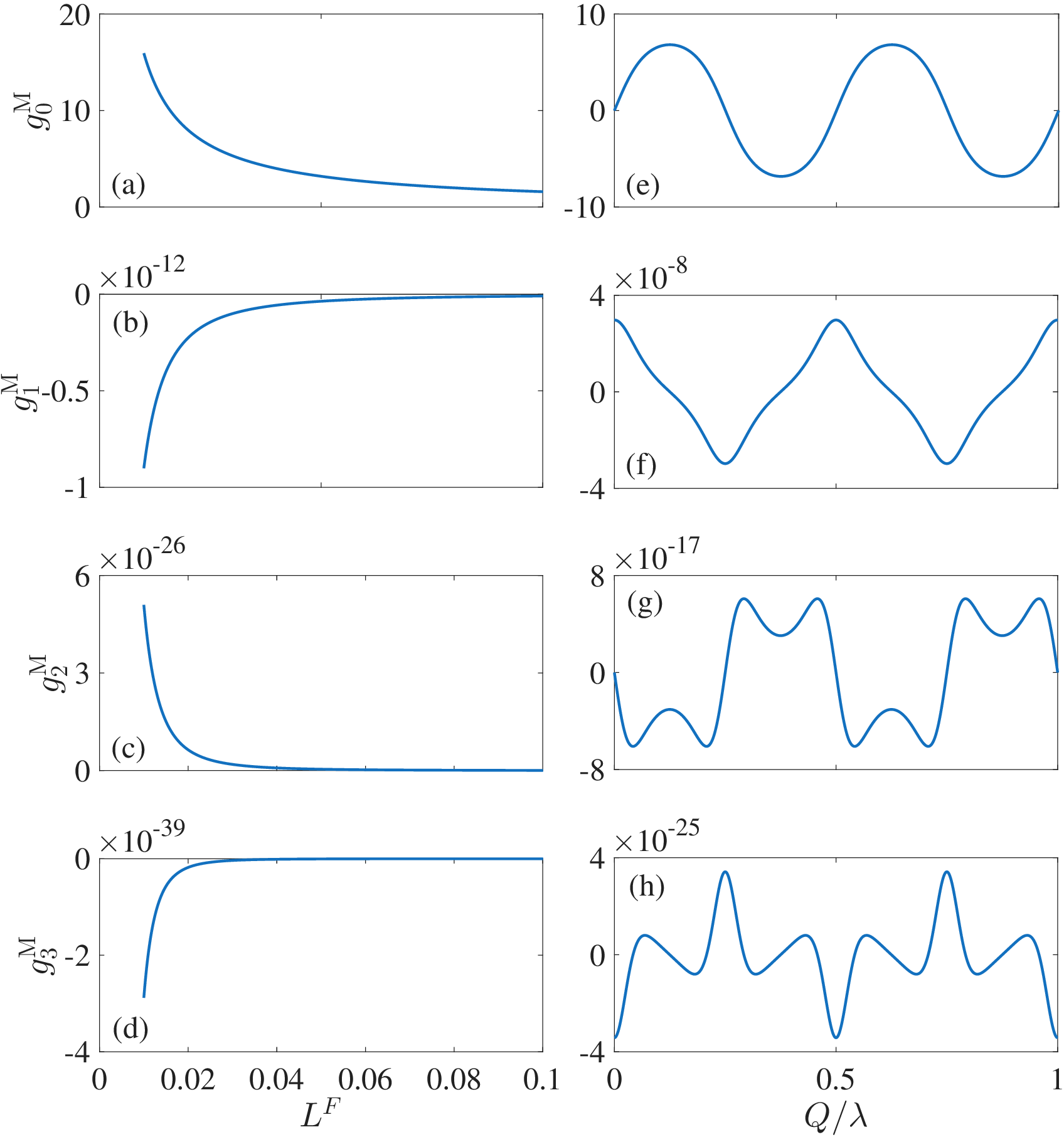}  
\caption{(a)–(d): Dependence of the first- to fourth-order optomechanical coupling coefficients of the FP OMS on the cavity length $L^F$. (e)–(h): Dependence of the first- to fourth-order optomechanical coupling coefficients of the MiM OMS on the membrane position $Q$ for a fixed cavity length. Parameters are taken from Tab.~\ref{tab}. 
\label{fig:3}}
\end{figure}

Figure~\ref{fig:3} shows the first four orders of the optomechanical coupling coefficients for a FP OMS and a MiM OMS. For a fixed mechanical resonator mass, the coupling strength in an FP OMS depends only on the cavity length $L$. In contrast, the coupling in a MiM OMS depends on both the cavity length $L$ and the position $Q$ of the membrane inside the cavity. Accordingly, the coupling coefficients in Figs.~\ref{fig:3}(a)-(d) are plotted versus cavity length $L$, while those in Figs.~\ref{fig:3}(e)-(h) are shown as functions of the membrane position $Q$ at fixed cavity length.

As shown in Fig.~\ref{fig:3}, the coupling coefficients of the FP OMS exhibit a monotonic dependence on cavity length $L^{\rm F}$ at every order, and all orders share the same sign. The ratios of successive couplings further satisfy the scaling relation $g_{n-1}/g_n \simeq L^{\rm F} x_{\rm eff}  $. In contrast, the MIM OMS displays a more complex behavior: the coupling strength varies non-monotonically with membrane position $Q$, and couplings of same orders can have opposite signs. At certain positions within the cavity, the coupling strength of specific orders vanishes. To enable direct comparison with values reported in the literature, we set the first-order coupling strength to $5$\,Hz. For a FP OMS with cavity length $L^{\rm F}=0.02$\,m, the second-, third-, and fourth-order couplings then fall on the order of $10^{-12}$\,Hz, $10^{-26}$\,Hz, and $10^{-39}$\,Hz, respectively. For the MiM OMS, four positions within one period satisfy the requirements: $Q_{1,2,3,4}/\lambda= 0.055$, $0.194$, $0.305$ and $0.445$. At these positions the first-order coupling takes the values $+5$\,Hz, $+5$\,Hz, $-5$\,Hz, and $-5$\,Hz, while the second- through fourth-order couplings reach magnitudes of approximately $10^{-8}$\,Hz, $10^{-17}$\,Hz, and $10^{-25}$\,Hz, respectively, with signs that depend on both the order and the chosen position.

\begin{figure}[]
\centering
\includegraphics[width=3.4in]{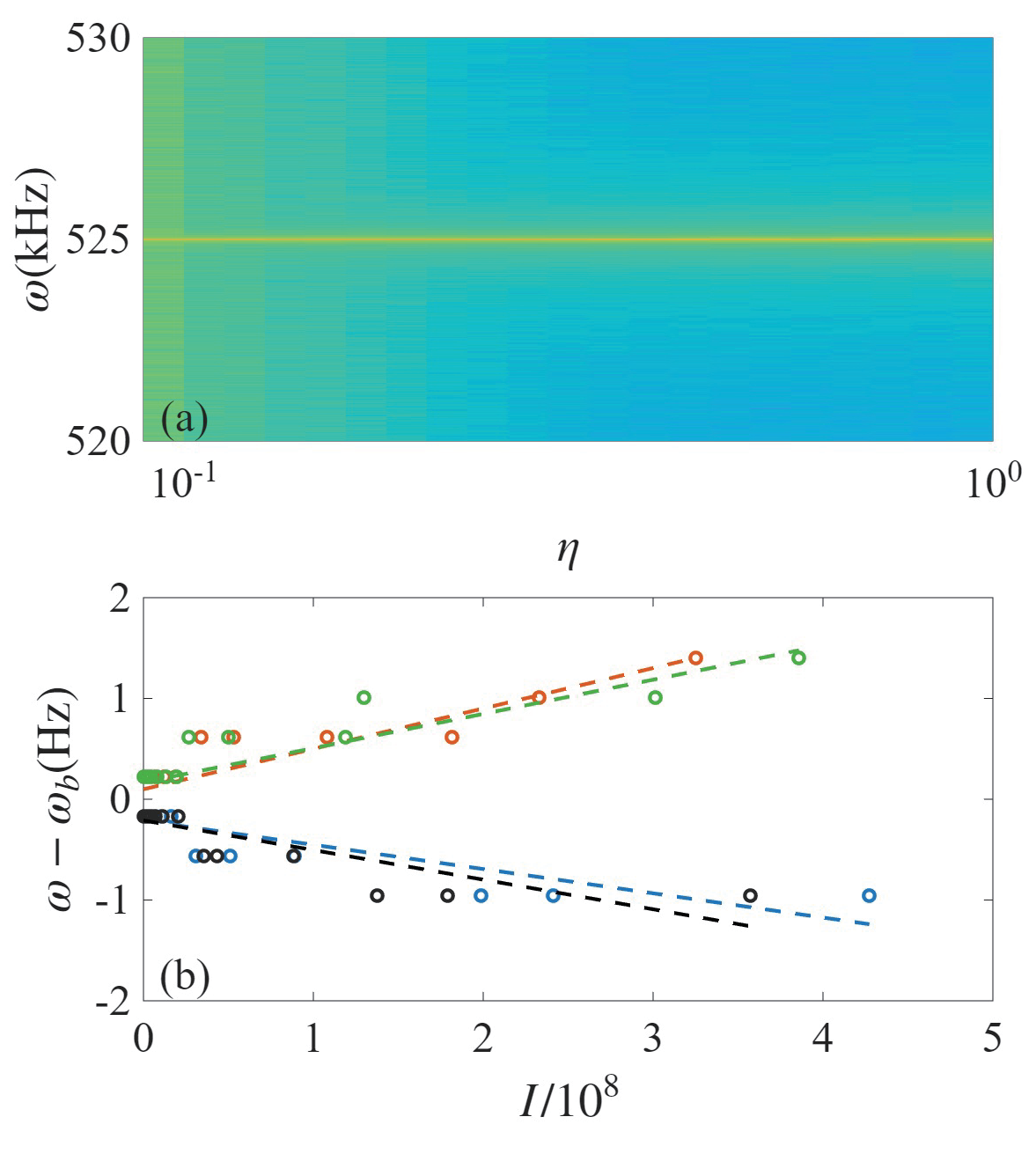}  
\caption{(a): Output spectrum of the probe field as a function of the power factor $\eta$ when the membrane is positioned at $Q_1/\lambda=0.055$. (b): Peak frequency of the probe spectrum versus oscillator amplitude for the membrane positions $Q_1/\lambda=0.055$ (blue), $Q_2/\lambda=0.194$ (red), $Q_3/\lambda=0.305$ (green), and $Q_4/\lambda=0.445$ (black). Dashed lines of corresponding colors represent linear fits to the data. Here we set $\phi=0$ to eliminate the influence of the calibration signal. Other parameters are the same as those used in Fig.~\ref{fig:2}.
\label{fig:4}}
\end{figure}
\begin{figure}[]
\centering
\includegraphics[width=3in]{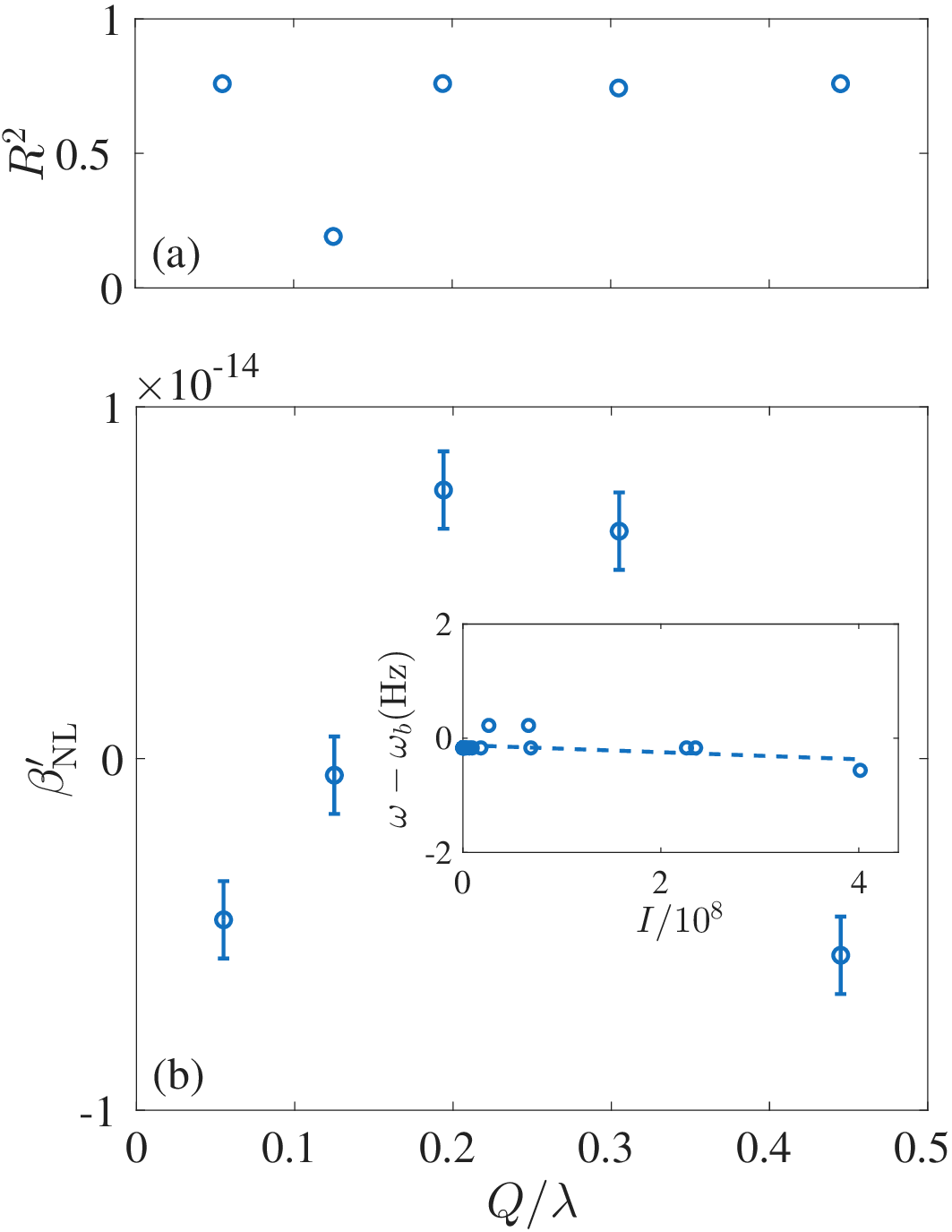}  
\caption{(a) Linear regression coefficients $R^2$ corresponding to the measured spurious signal $\beta'_{\rm NL}$ for different membrane positions. (b) Magnitude of the spurious $\beta'_{\rm NL}$ together with its $95\%$ confidence interval. Then inset in (b): Peak frequency of the probe spectrum versus oscillator amplitude for the membrane positions $Q_5/\lambda=0.125$. Dashed line represents linear fit to the data. The parameters are the same as those used in Fig.~\ref{fig:4}.
\label{fig:5}}
\end{figure}

Given the more intricate dynamics and stronger higher-order nonlinearities in the MiM OMS, we perform numerical simulations of the measurement process for this system to illustrate the influence of these higher-order terms on the proposed GUP measurement scheme. Fig.~\ref{fig:4} presents the output spectra obtained by numerically solving Eq.~\eqref{eq:MB_CLE} with $\gamma t_p=5.25\times 10^{-4}$ and $\gamma\Delta t=2.625\times 10^{-3}$, following the procedure described above. We first consider the case in which the driving fields $E_1$, $E_2$ and the probe field $E_p$ are modulated simultaneously. We define a power factor $\eta$ and set the driving powers to $P_1=\eta^2\times 2\,\mu\mathrm{W}$, $P_2=\eta^2 \times 50\,\mu\mathrm{W}$, and $P_p=\eta^2\times 5\,\mu\mathrm{W}$. Here we choose $\beta_{\rm NL}=0$, under this ideal condition the oscillator’s  frequency remains constant and the spectrum therefore should be independent of the laser intensity. Fig.~\ref{fig:4}(a) shows the output spectrum as a function of the power factor $\eta$ when the membrane is placed at position $Q_1$. Following the data processing procedure, we extract the peak frequencies and plot them against the corresponding mechanical oscillation amplitudes $I$ in Fig.~\ref{fig:4}(b) (blue scatter points). Although $\beta_{\rm NL}=0$, the frequency of the probe peak still shifts with oscillator amplitude, thereby generating a spurious signal $\beta'_{\rm NL}$. A linear fit to the data (blue dashed line) yields $\beta'_{\rm NL}\approx -4.59\times 10^{-15}$, a value many orders of magnitude larger than the genuine GUP signal. The red, green, and black scatter points in Fig.~\ref{fig:4}(b) correspond to the membrane positions $Q_2$, $Q_3$, and $Q_4$, respectively; the dashed lines of matching colors represent the corresponding fits. The resulting spurious values are $\beta'_{\rm NL} \approx 7.64\times 10^{-15}$, $\beta'_{\rm NL} \approx 6.47\times 10^{-15}$, and $\beta'_{\rm NL} \approx -5.59\times 10^{-15}$. The signs of these spurious signals are consistent with the quadratic coupling behavior shown in Fig.~\ref{fig:3}(b). We plot these results, together with the corresponding regression coefficients $R^2$ that quantify the reliability of the linear fits, in Fig.~\ref{fig:5}. The obtained $R^2$ values are approximately $0.8$. This indicates that if the origin of $\beta'_{\rm NL}$ in the higher-order nonlinearities of the OMS were not recognized, the spurious signal could be mistaken for a highly reliable GUP signature.

It is worth noting that the influence of the second-order coupling on the measurement can be mitigated by adjusting the membrane position. Here we place the membrane at $Q_5/\lambda=0.125$, which, as shown in Fig.~\ref{fig:3}(b), corresponds to a local extremum of the first-order coupling where the second-order coupling is approximately zero. Experimentally, such an extremum can be identified by continuously scanning the membrane position while monitoring the first-order coupling coefficient. The inset in Fig.~\ref{fig:5}(b) presents the simulated measurement results obtained with the membrane at this position. The peak frequency remains essentially independent of the oscillation amplitude, indicating that the spurious signal is effectively suppressed. Moreover, Fig.~\ref{fig:5}(a) shows that the corresponding $R^2\simeq 0.1$. Under these conditions, previous studies would regard the signal as falling below the resolution limit, implying that no GUP effect was detected. It can therefore be concluded that, in this case, higher-order OMS effects did not give rise to a spurious signal.

\begin{figure}[]
\centering
\includegraphics[width=3in]{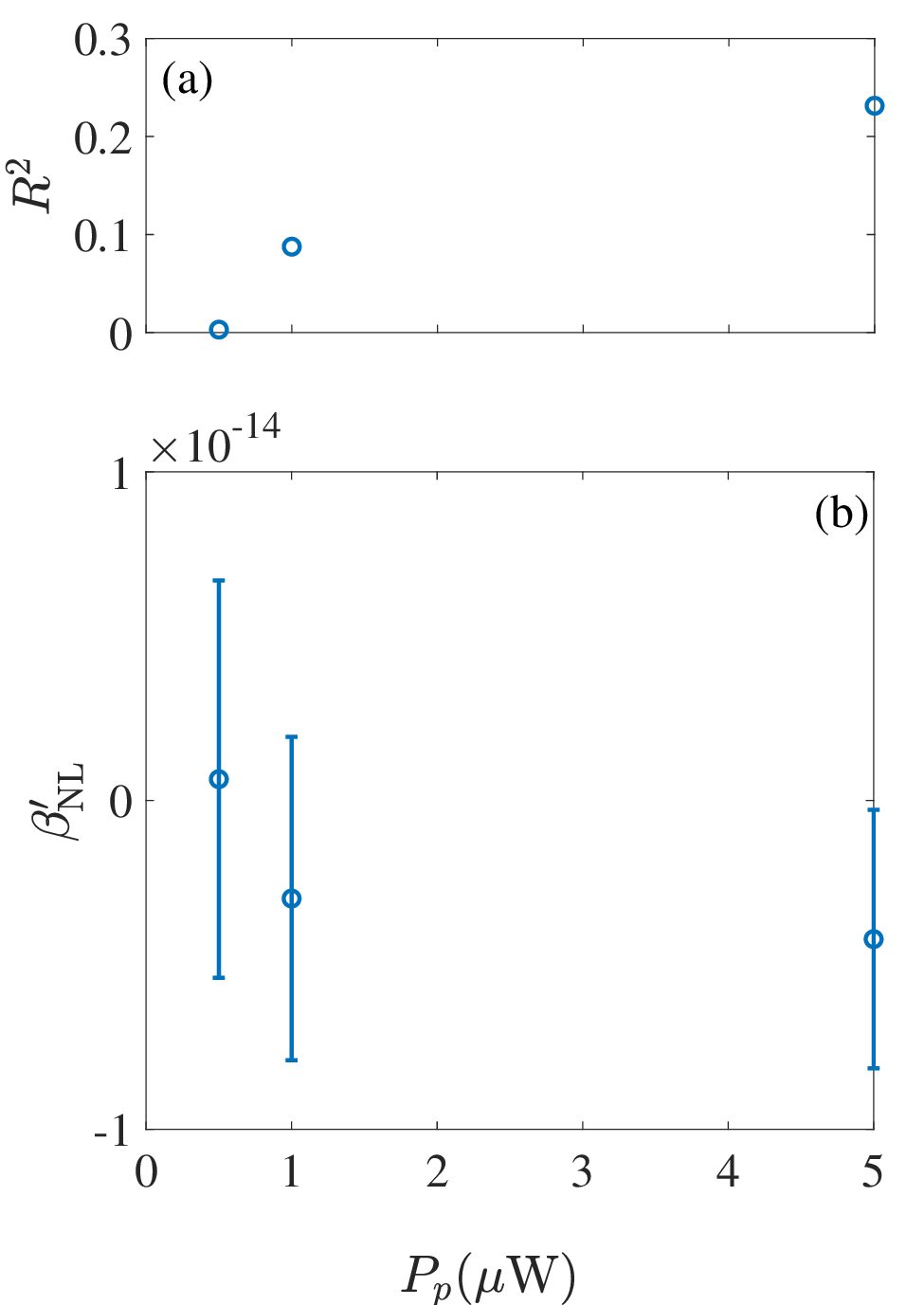}  
\caption{(a): Linear regression coefficients $R^2$ corresponding to the measured spurious signal $\beta'_{\rm NL}$ for different input powers of the probe field. (b): Magnitude of the spurious $\beta'_{\rm NL}$ together with its $95\%$ confidence interval. The membrane is positioned at $Q_1/\lambda=0.055$, with  $P_2=50\mu$W fixed.$P_1$ is adjusted according to $P_1=\eta^2\times 2\,\mu\mathrm{W}$, where $\eta$ takes $19$ discrete values uniformly spaced from $0.01$ to $0.1$. Other parameters are the same as those used in Fig.~\ref{fig:4}.
\label{fig:6}}
\end{figure}

If the probe field is held constant while the driving-field intensity is varied, the last term in Eq.~(41) becomes a constant. Consequently, the slope of the linear relation between the effective frequency and the squared mechanical amplitude transforms according to:
\begin{equation}
\begin{split}
\beta_{\rm NL}\rightarrow \beta_{\rm NL}-12g_3\vert a_2\vert^2,
\end{split}
\label{eq:omega_eff_full_a2}
\end{equation}
It can be seen that the magnitude of the spurious signal depends on the fourth-order coupling strength $g_3$ and is amplified by the probe-field intensity $\vert a_2\vert^2$. Typically, the driving power of the probe field lies in the microwatt range, corresponding to an intracavity intensity of order $\vert a_2\vert^2\sim 10^9$ for the parameters used here. Consequently, for an FP OMS, $g_3$ is of order $  10^{-39}$ and the resulting spurious contribution is approximately $\beta'_{\rm NL}\sim 10^{-30}$. For an MiM OMS, $g_3$ is of order $ 10^{-25}$ and the spurious signal reaches approximately $\beta'_{\rm NL}\sim 10^{-15}$, a level comparable to the resolution limits of existing OMS-based measurement schemes reported in Refs.~\cite{Li2025,Li2026,Li20262}. As shown in Fig.~\ref{fig:6}, when the input power of the probe field reaches $5\mu$W, a nonlinear signal with an amplitude of $\beta'_{\rm NL}\sim 4\times10^{-15}$ is measured. This signal, arising from the fourth-order coupling $g_3$, is indistinguishable from the GUP effect and corresponds to an $R^2$ of $0.3$. As the probe-field power is reduced to $0.5\mu$W, both the magnitude of the signal $\beta'_{\rm NL}$ and the corresponding $R^2$ approach zero. This implies that the probe field should be kept as weak as possible in order to minimize the spurious contribution. However, in the presence of phase noise the probe cannot be made arbitrarily weak, otherwise the signal would be overwhelmed by noise. A quantitative analysis is therefore required for each experiment to determine the optimal probe intensity and the associated magnitude of the residual spurious signal.

\subsection{Impact of driving field phase noise on quantum regime measurements}

As discussed previously, phase noise introduces additional noise into the input spectrum but does not overwhelm the signal when the oscillator remains in the classical regime. This indicates that phase noise does not fundamentally invalidate the measurement scheme for classical harmonic oscillators. However, when a high-purity quantum harmonic oscillator is used as the probe, as considered in Ref.~\cite{Li2025}, phase noise in the driving field degrades laser cooling and thereby limits the achievable purity of the oscillator. During the subsequent measurement phase, the cooling and pump fields are switched off. Under the condition $  \beta_{\rm NL}\ll 1  $, the oscillator then evolves under free dissipation and thermalization with the thermal bath, so that its effective phonon number follows~\cite{Liu2013}:
\begin{equation}
\begin{split}
\bar{n}(t)\simeq \bar{n}(0)e^{-2\gamma t}+\left(1-e^{-2\gamma t}\right)\bar{n}_b.
\end{split}
\label{eq:phonon number}
\end{equation}
Here, $\bar{n} =\sqrt{\langle \delta^2 q\rangle\langle \delta^2 p\rangle-\langle \delta q\delta p\rangle^2}$ denotes the effective phonon number of a squeezed Gaussian state, with $\langle\delta^2 o\rangle=\langle o^2\rangle-\langle o\rangle^2$ and $\langle\delta o_1 o_2\rangle=\langle o_1 o_2\rangle-\langle o_1\rangle\langle o_2\rangle$. The purity of the quantum state is given by $\mathcal{P}=1/(2\bar{n}+1)$~\cite{Paris2003}. Its time evolution during free thermalization reads:
\begin{equation}
\begin{split}
\mathcal{P}(t)\simeq \dfrac{\mathcal{P}(0)}{e^{-2\gamma t}+\dfrac{\mathcal{P}(0)}{\mathcal{P}_n}\left(1-e^{-2\gamma t}\right)},
\end{split}
\label{eq:phonon number}
\end{equation}
where $\mathcal{P}_n$ is the purity corresponding to thermal equilibrium with the environment and $\mathcal{P}(0)$ is the initial purity at the beginning of the measurement. Consequently, a lower initial purity causes the oscillator to thermalize into a classical state more rapidly. If a minimum purity threshold (e.g., $\mathcal{P}>0.01$) is imposed to maintain quantum behavior, the available measurement time window is shortened. A shorter time window in the time domain directly degrades the frequency resolution and therefore compromises the overall measurement precision.

\begin{figure}[]
\centering
\includegraphics[width=3in]{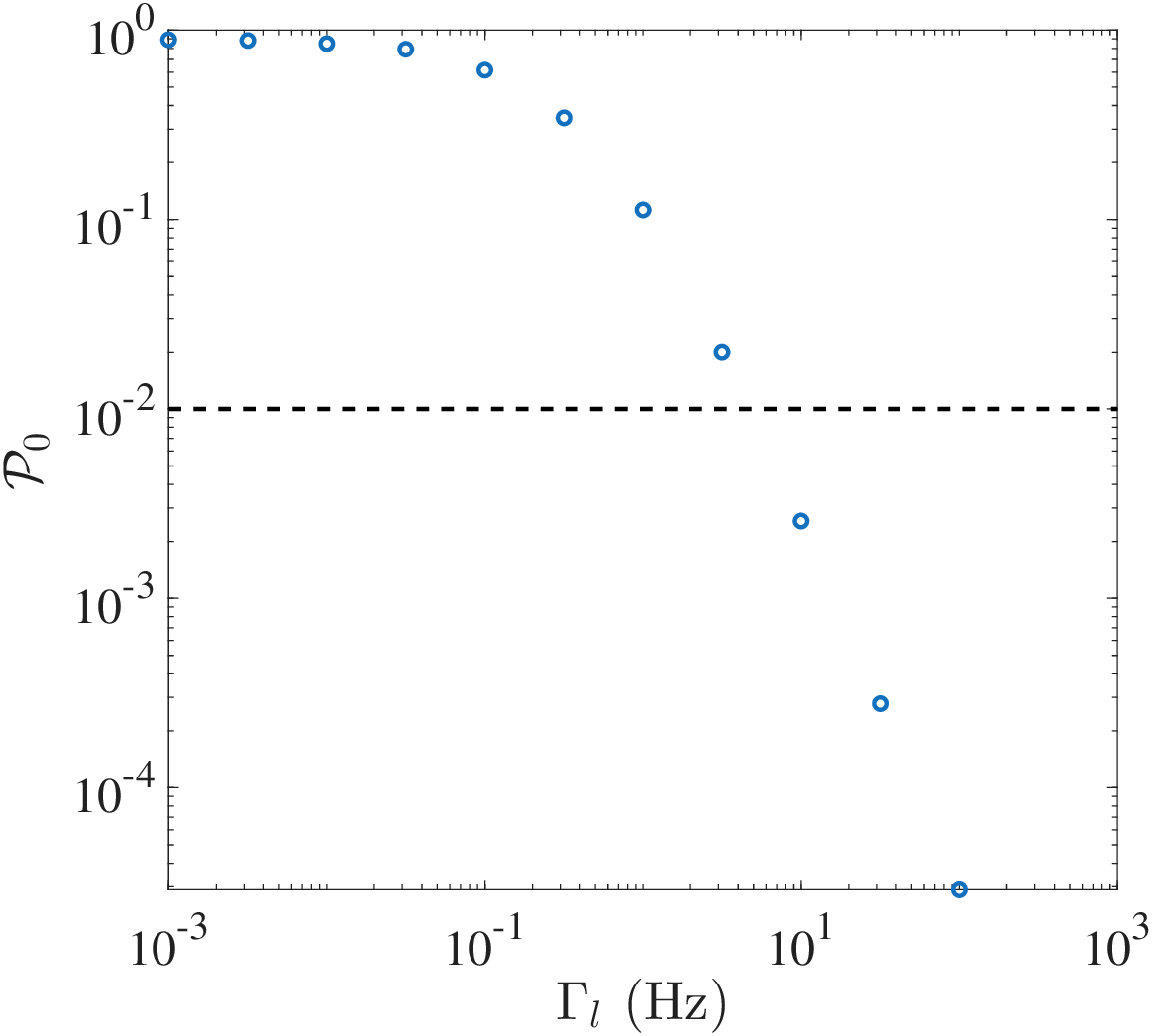}  
\caption{Oscillator purity at time $t_p$ (the instant at which the cooling and pump fields are switched off and data acquisition begins) as a function of the phase-noise amplitude. The covariance matrix used to evaluate the purity is obtained from $25,000$ independent realizations of the stochastic Langevin equation Eq.~\eqref{eq:MB_CLE}. Here, the membrane is placed at $Q_1$ and we set $\Delta_2=\omega_b$. The other parameters are the same as those in Fig.~\ref{fig:2}.
\label{fig:7}}
\end{figure}
Figure~\ref{fig:7} illustrates the purity achieved by the oscillator after the cooling process for different phase-noise amplitudes. A reduction in the initial purity leads to a significant degradation of the measurement resolution. According to Eq.~\eqref{eq:phonon number}, when the ambient temperature is set to $0.1$\,K, the oscillator’s equilibrium phonon number reaches the order of $\bar{n}_b\sim 10^3$, corresponding to a steady-state purity of order $\mathcal{P}_n\sim 10^{-4}$. Under these conditions, a rate of only $\Gamma_l\sim 1$\,Hz is sufficient to double the minimum resolvable frequency (i.e., to degrade the resolution). At still higher phase-noise amplitudes, measurements based on the quantum oscillator become infeasible under the present parameters unless the cooling-field parameters are further optimized.

\section{Discussion and Conclusion}
\label{Discussion of other situations}
In summary, we have reexamined OMS employed as detectors for quantum gravity effects using a more complete dynamical treatment. Our analysis demonstrates that the conventional idealized description, which retains only the first-order radiation-pressure interaction, is insufficient. Higher-order optomechanical couplings, although individually weak, can generate spurious signals comparable in magnitude to the minute quantum gravity corrections they are intended to detect. Furthermore, excess laser noise, particularly phase noise, adversely affects the noise profile of the measurement scheme. Taking the detection of GUP effects through the dynamics of a harmonic oscillator as a concrete example, we have analyzed two representative OMS platforms, the FP OMS and MiM OMS, incorporating the full nonlinear dynamics together with realistic laser phase noise. The principal conclusions that bear directly on experimental practice are as follows: (1). When the driving-field intensity is varied, the probe-field intensity must be held strictly constant; otherwise, the second-order coupling induces an amplitude-dependent shift in the effective frequency, producing a false signal. (2). The probe power should be kept as low as possible while still ensuring that phase noise does not dominate the signal; otherwise, the intracavity photon number amplifies the frequency shift arising from fourth-order coupling, again generating a spurious contribution. (3). Precision measurements of a quantum harmonic oscillator require effective suppression of laser phase noise. (4). In MiM OMS, the membrane should be positioned at an extremum of the first-order coupling coefficient. This optimal working point can be located experimentally by continuously scanning the membrane position while monitoring the signal.

By evaluating realistic experimental parameters, we find that in typical FP systems the second- and fourth-order coupling strengths are on the order of $10^{-12}$\,Hz and $10^{-39} $\,Hz, respectively, for intracavity photon numbers of order $10^9$. The resulting perturbations to the extracted value of $\beta_{\rm NL}$ are therefore approximately $10^{-19} $ and $10^{-30}$. In MIM systems the corresponding couplings reach $10^{-8}$ Hz and $10^{-25}$ Hz, leading to perturbations of order $10^{-15}$ and $10^{-15}$, respectively. These values substantially larger than in the FP case and therefore more consequential for precision measurements.

Our study has focused on the standard optomechanical scheme for GUP detection employed in Ref.~\cite{Li2025}. The influence of full system dynamics is expected to be even more significant in modified protocols, such as dark-mode interference~\cite{Li2026} or higher-order sideband techniques~\cite{Li20262}. Moreover, the conclusions drawn here are not restricted to GUP measurements; they apply equally to other weak-signal optomechanical detection schemes, including those aimed at Schrödinger–Newton gravity~\cite{Yang2013,Grossardt2016,Tang2025,Yan2025}, gravitational interaction~\cite{Bonaldi2026}, quantum collapse model~\cite{Li2016} or quantum illumination~\cite{Barzanjeh2015}. Ultimately, high-precision metrology rests upon a correspondingly rigorous characterization of the measurement apparatus itself. 

Another noteworthy factor is the weak nonlinearity inherent to the mechanical oscillator itself, arising from material structure, defects, and asymmetry~\cite{Bawaj2015,Serra2016,Shen2022}. This nonlinearity appears in the Hamiltonian as a Kerr-type term involving higher-order powers of the position coordinate. Although it is distinguishable from the momentum-dependent Kerr nonlinearity associated with the GUP, its quantitative contribution must still be measured and carefully accounted for in the experimental analysis. Furthermore, because we are concerned with the center-of-mass degrees of freedom of the oscillator, the effective GUP parameter may undergo a material dependent renormalization of the form $\beta_0/N^s$, where $s>0$ is an unknown exponent related to the intrinsic properties of the material and $N$ denotes the number of fundamental GUP units~\cite{AmelinoCamelia2013,Kumar2020,Bosso2023}. These issues remain to be investigated in future work.

\begin{acknowledgements}
W.L. is supported by the National Natural Science Foundation of China (Grants No.~12304389), the Scientific Research Foundation of NEU (Grant No. 01270021920501*115). Yan~Li is supported by the Fundamental Research Funds for the Central Universities (Grant No. GK202601011). C.Z. is supported by the National Natural Science Foundation of China (Grant No. 12447152). 
\end{acknowledgements}

\section*{DATA AVAILABILITY}
The data that support the findings of this article are openly available~\cite{data}; embargo periods may apply.

\end{document}